\documentclass{elsarticle}
\usepackage{graphicx}
\usepackage{subfig}

\usepackage[colorlinks=true,urlcolor=blue]{hyperref}

\begin{document}
\begin{frontmatter}

\title{Characterization of an electrically cooled BEGe detector till E$_{\gamma}$$\sim$7 MeV}

\author[label1]{Sathi Sharma} 
\author[label2]{Arkabrata Gupta} 
\author[label4]{Balaram Dey} 
\author[label3]{M. Roy Chowdhury} 
\author[label3]{A. Mandal} 
\author[label2]{A. Bisoi} 
\author[label3]{V. Nanal} 
\author[label3]{L.C. Tribedi} 
\author[label1]{M. Saha Sarkar \corref{cor}}
\cortext[cor]{Corresponding author.}
\ead{maitrayee.sahasarkar${@}$saha.ac.in}

\address[label1]{Saha Institute of Nuclear Physics, HBNI, 1/AF Bidhannagar, Kolkata - 700 064, India}
\address[label2]{Indian Institute of Engineering Science and Technology, Shibpur, Howrah - 711 103, India}
\address[label3]{Department of Nuclear and Atomic Physics, Tata Institute of Fundamental Research, Colaba, Mumbai - 400 005, India}
\address[label4]{Bankura University, Bankura, West Bengal - 722 155, India}

\begin{abstract}

An electrically cooled Broad Energy Germanium (BEGe) detector has been characterized in the energy range E$_{\gamma}$ $\sim$ 0.122 - 7 MeV by utilizing the $\gamma$- rays emitted by a short-lived  resonance state in $^{15}$O populated through $^{14}$N(p,$\gamma$) reaction  and standard  radioactive source ($^{152}$Eu). The experimental results have been reproduced through simulations with GEANT4 code, including vendor specified detector geometry along with the detailed construction of the target holder flange,  to delineate the effects of the holder at various energies and detector position. 
Later the efficiency with a bare point source has been simulated. It has been found that the electrically cooled BEGe detector is suitable for usage in the $\gamma$- ray spectroscopy as well as for the study of resonance phenomena in nuclear astrophysics. %till 7 MeV.  
\end{abstract}\end{frontmatter}

%\keywords{{\it Keywords}:  Gamma  detectors (BEGe, HPGe); Electrically cooled detector; Resonance reaction; Efficiency; Simulation of detector response; High energy gammas}                                                                                                                             

% main text
\section{Introduction}
Nuclei lighter than calcium, usually possess de-excitation $\gamma$- rays of high energy ($>$ 3  MeV). Thus, $\gamma$- detectors used in low energy Nuclear Astrophysics experiments planned to study light element nucleosynthesis are characterized for these energies. Most of these reactions at stellar energies have very low cross-section. The response of the detectors in terms of full energy peak detection efficiencies, the relative intensities of escape peaks for high energy gammas, their contribution in the low energy, and Compton continuum background are needed to identify and estimate the intensities of $\gamma$- rays arising from these reactions. 

Radioactive sources  generally used in the laboratories (like $^{152}$Eu, $^{60}$Co, $^{133}$Ba, $^{207}$Bi etc.) emit $\gamma$- rays up to 1.7 MeV. So the higher energy ($>$ 2 MeV) characteristics of $\gamma$- detectors are either determined from simulation or extrapolation of the empirical data taken at lower energies. Proton capture resonance reactions in lighter nuclei like $^{13}$C, $^{14}$N, $^{15}$N produce high energy $\gamma$- rays as the Q value of these reactions are positive and quite high ($\simeq$ 7 MeV or more). So $\gamma$- rays emitted by these resonance states have 
been utilized to characterize $\gamma$- detector efficiencies till high energies ($\simeq$ 11 MeV \cite{clover}).

High Purity Germanium (HPGe) detectors are invaluable in nuclear physics research due to their excellent energy resolution.
They are operated in a specially-designed cryostat cooled with liquid nitrogen (LN$_2$) to reduce thermal noise. However, in recent times, different techniques have been developed to replace LN$_2$ cooling with mechanical or electrical cooling without affecting the performance of the detector - especially energy resolution.

\begin{figure}[h]
\centering
\includegraphics[angle=270,width=1.0 \textwidth]{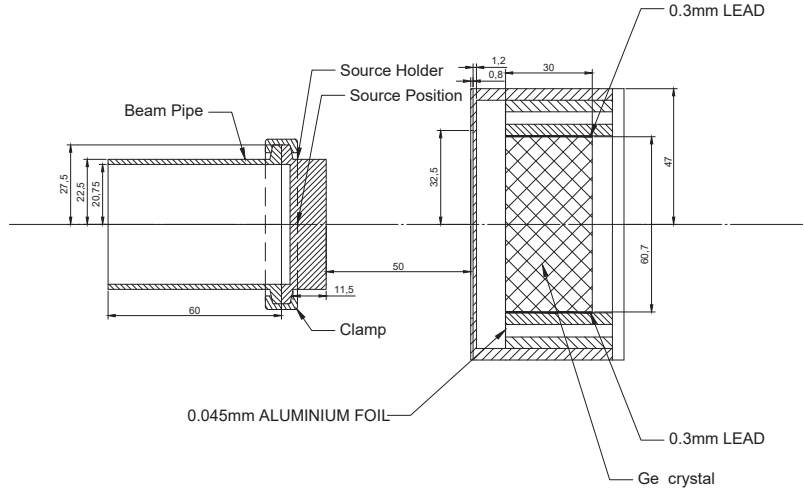}
\caption{
\label{fig1_falcon}
 The internal structure for holding the Ge crystal: as provided by the manufacturer for the BEGe detector, source holder, and beam pipe geometry, are shown in this figure. These specifications are used in the simulation.}
\end{figure}

\begin{figure}[h]
\begin{center}
\includegraphics[width=\textwidth]{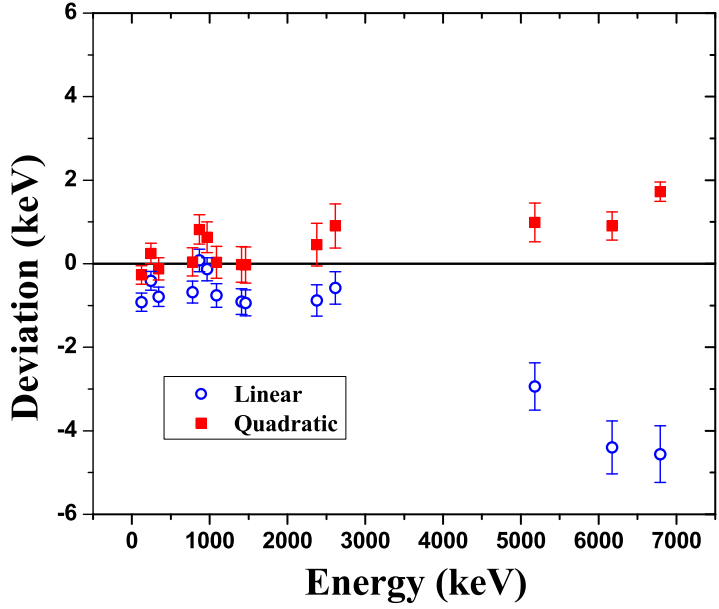}
\caption{\label{fig2_calib} \small \sl Deviations (keV) from the linear and quadratic fit to the calibration data points of the Falcon BEGe detector are plotted.  }
\end{center}
\end{figure}

\begin{figure}[h]
\begin{center}
\includegraphics[height=12.0cm,width=16.0cm]{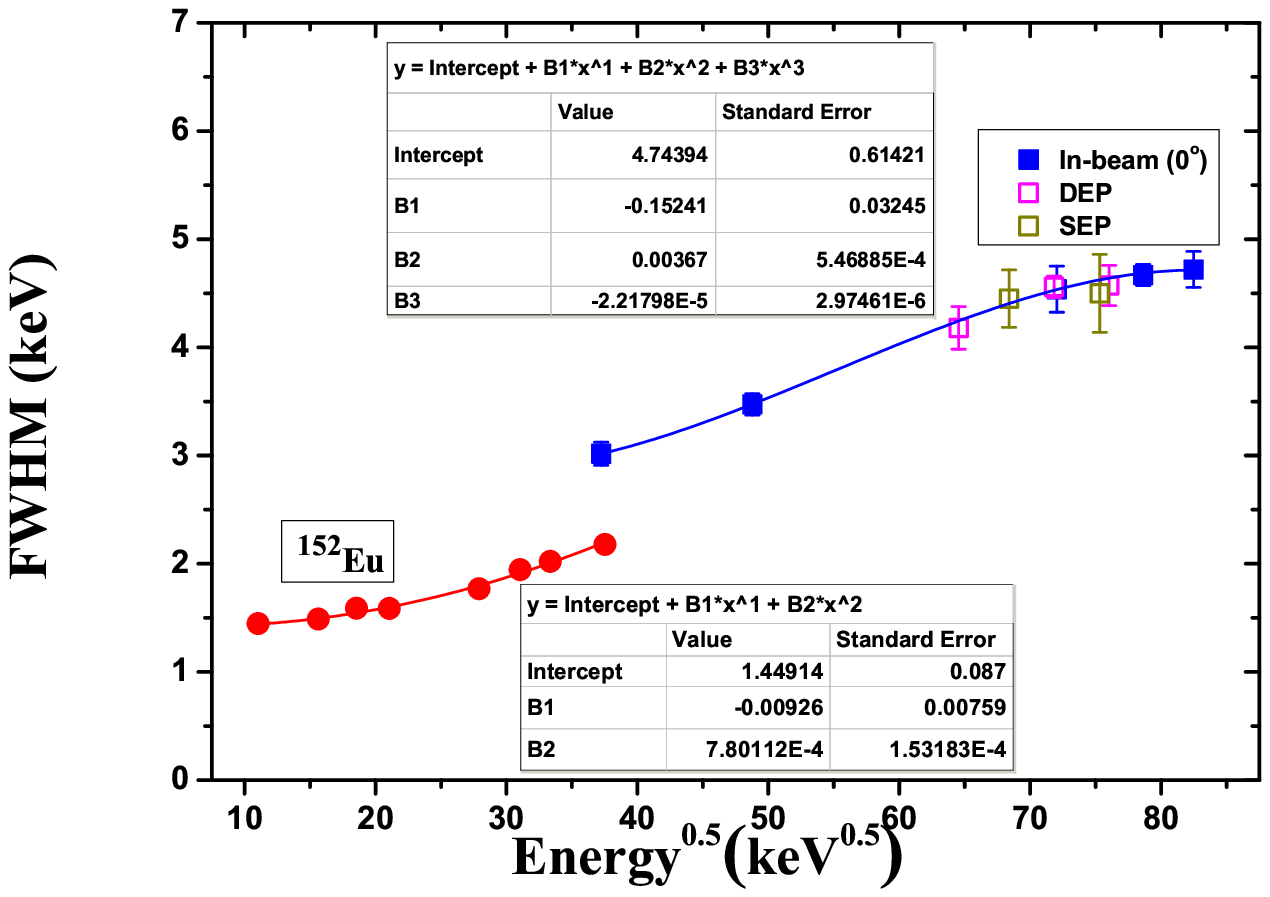}%{fig3_reso.eps}
\vspace{-3cm}
\caption{\label{fig3_fwhm} \small \sl Resolution (FWHM) vs. $\sqrt{E}$ plots for full energy peaks of (i)$^{152}$Eu source labeled with red circles (Set I) fitted with a quadratic function, 
(ii) in-beam $\gamma$- rays from the resonance state with detector at 0$^o$ with blue squares (Set II) fitted with a cubic function.
The FWHMs of single and double escape peaks of in-beam $\gamma$- rays from the resonance state with the detector at 0$^o$ are marked with dark yellow and magenta squares, respectively. The fitting parameters and their errors are tabulated in the in-set tables.
 }
\end{center}
\end{figure}

The Ge crystal of BEGe configuration offers a clear advantage over more conventional HPGes. The BEGe detectors \cite{bege} are designed with a special electrode structure to improve low energy resolution. The impurity doping in the germanium crystal is such that it also results in better high energy resolution and peak shape due to improvement in charge collection. These features enhance the efficiency and resolution at low energy while preserving good efficiency at higher energies. 

Several groups \cite{pulse, bege1, bege1a, bege2,bege3} have systematically investigated the performance of commercial BEGe detector
from energies below 10 keV \cite{bege1} (for a detector having mm thin carbon composite entrance window and $\mu$m thick front dead layer) till 1.5 MeV \cite{bege1a}. The Monte Carlo simulations for a large volume BEGe detector have been performed to calculate its detection efficiencies, which were validated over various realistic measurement scenarios \cite{bege1a}. BEGe detectors have discrimination power for distinguishing between multi-site and single-site energy deposition events \cite{bege2,bege3}. These detectors are currently studied extensively for their probable usage in double beta decay experiments \cite{bege2}.

The CANBERRA Broad Energy Ge (BEGe) Falcon 5000 detector \cite{falcon} is a commercially available electrically cooled detector. 
In the present work, a proton capture resonance reaction  $^{14}$N(p,$\gamma$) at E$_{p}^{lab}$= 278 keV has been utilized to populate a resonance state at around 
7.556 MeV in $^{15}$O. The resonance state decays by emitting $\gamma$- rays of energies till 7.556 MeV. The $\gamma$- rays emitted by this short-lived  resonance state  and a standard  radioactive source 
($^{152}$Eu) have been utilized to characterize the electrically cooled BEGe detector in the $\gamma$- energy 
range E$_{\gamma}$ $\sim$ 0.122 - 7 MeV. The response of the detector has also been simulated using GEANT4.10.00.p04 package \cite{geant4} and compared with the experimental data.\\

\begin{table}[h]
\caption{Specifications of the BEGe detector used in the present work.}
\label{detector}
\begin{center}
\begin{tabular}{ccc|cc|cc}
\hline
\multispan{3}\hfil  Detector \hfil&\multispan{2}\hfil  Ge Crystal \hfil&
\multispan{2} \hfil  Al window\hfil\\
&&&\multispan{2}\hfil  (mm)\hfil&\multispan{2} \hfil  (mm)\hfil \\
\multispan{3}\hrulefill & \multispan{2}\hrulefill &
\multispan{2} \hrulefill
\\
Make&Type&High&Dia.& Height&  Dist.\footnote{1}& Thick\\
        &&Voltage(V)&& & 
& \\ \hline
Canberra&Falcon 5000&-3700&60&30&12&0.12\\
(Mirion)     &BEGe&&&&&\\
   &&&&&&\\
%\strut \\
%BSI\footnote{2}&p-type&&&&& \\
%&GCD-30 185&+2600&54&63&6&0.7\\
%%  &&  & \\
\hline
\end{tabular}\end{center}
{$^1$}{Distance between the crystal and the endcap.} \\
%{$^2$}Baltic Scientific Instrument. 
\end{table}

\begin{table}[h]
\caption{Specifications of the source holder and associated parts. Please see Fig. \ref{fig1_falcon} for nomenclature.}
\label{holder}
\begin{center}
\begin{tabular}{c|c|c|c}
\hline
Item Name& \multispan{3}\hfil  Detail Specifications (cm)\hfil\\
&\multispan{3} \hrulefill
\\
& Length & \multispan{2}\hfil Diameter\hfil\\
&&\multispan{2} \hrulefill
\\
&&Inner&Outer\\
\hline
Source/Target holder & 1.150 & 0.0 & 2.250 \\
Clamp & 0.600 & 2.750  &  3.250 \\
Beam pipe & 6.000 & 2.075 & 2.250 \\
\hline
\end{tabular}
\end{center}
\end{table}

\section{Experimental details and Simulations}
\subsection{Experimental set up, detectors, data acquisition and analysis} 

The present experiment was performed at TIFR, Mumbai, using ECR (Electron Cyclotron Resonance) ion-source based low energy ion accelerator \cite{lokesh1,lokesh2,lokesh3}. 
An implanted $^{14}$N target previously characterized \cite{abhijit} was bombarded by proton beam at lab energy E$_{p}^{lab}$= 278 keV. The typical beam current was 3 - 4 $\mu$A. 
The target was mounted at the end-flange at 0$^o$ beamline. A schematic diagram of the experimental set-up is shown in Fig. \ref{fig1_falcon}. 

A FALCON BEGe detector was placed at different angles, {\it viz.} 0$^o$, 50$^o$ and  90$^o$ with respect  to the beamline at a distance of 5 cm 
from the end flange centre (target centre).  

The FALCON detector is a complete stand-alone system with a full version of data acquisition and analysis GENIE-2000 \cite{genie} software included. The unit also includes a High Voltage Power Supply (HVPS)  for the detector and Digital Signal Processing  (DSP) unit. A reverse bias of -3700 V, provided by the HVPS unit is applied to the detector. The DSP unit takes care of the input pulse processing and amplification. GENIE software has a comprehensive set of capabilities for acquiring and analyzing spectra from Multichannel Analyzers (MCAs). The software can handle several functions like,  MCA control, spectral display and manipulation, basic spectrum analysis and reporting 
\cite{genie}. The amplifier gain was kept to a minimum value and the spectrum was chosen to have a maximum  number of channels (8k). It was ensured that high energy 
$\gamma$- rays up to 8 MeV can be seen in the spectrum. The primary acquisition and analysis  were  done with the GENIE software. 
However,  for final analysis, an offline analysis programme INGASORT \cite{Ranjan} was utilized. 

Low energy $\gamma$- rays were attenuated substantially, even at the 0$^o$ position of the detector as the thick stainless steel (SS) end--flange of the beamline was used as the target holder. 
Attenuation varied at different angular positions of the detector due to the specific design of the end--flange. Thus, the efficiency with standard $^{152}$Eu source placed at the target position has been determined experimentally for all positions of the detector. A calibrated
$^{152}$Eu (35298 (82) Bq measured on 1$^{st}$ June, 2011) radioactive source has been used for low energy calibration and $\gamma$- efficiency (E$_{\gamma}$ = 0.122 - 1.408 MeV)  response of the detector. 

However, for experimental efficiency calibration at higher energies with $\gamma$- rays emitted from in-beam resonance, the beam-spot was found to show a shift from the centre of the target (the position of the $^{152}$Eu source during off-beam measurements). Thus, to match the $^{152}$Eu and in-beam resonance data for determination of the efficiency calibration of the detector over the whole energy range,  simulation has been done. The simulated data have been first matched with the experimental data, including the exact experimental conditions (the target position, the target holder geometry, etc.). Later absolute efficiency of the detector is simulated for point sources. Thus,  the effects of target holder asymmetries as well as source position mismatch were
eliminated.  

The measurements reported in this paper were acquired in a singles mode for  sufficient time  to keep statistical error minimum.
The contributions from room background were properly taken care of and the beam-induced $\gamma$- ray background contributions were monitored with beam-on through a blank target frame.

\subsection{Essential details of GEANT4 simulation } 

A Monte-Carlo GEANT4 \cite{geant4} simulation has been performed in order to study the response of the detector. It has been useful to understand the
effect of the target holder geometry in the variation of efficiencies at different angles. 

GEANT4 is designed to simulate the passage of particles through matter and comes pre-configured with many standard options and examples for a user to modify. The simulation was done using a series of GEANT4 classes like
detector construction and material building, particle and physics process definition, particle tracking, event action, etc.
The geometry of the simulation was created as fully and accurately as possible considering all the beamline components such as end flange, beam pipe, clamp, etc. Individual $\gamma$- particles were randomly generated by a particle generator (G4ParticleGun) and tracked through the detector volume. 
Five to ten million events were thrown isotropically on the front face of the detector to
generate the simulated spectra of each $\gamma$- ray energy.  
The energy deposition was recorded step-by-step and finally added for each event. 
The processes of ionization, multiple scattering and bremsstrahlung were included for the electrons and positrons in the simulation code.
Similarly, the photoelectric effect, the Compton scattering and the
pair production were considered for $\gamma$- rays.

%In the simulation, low energy standard electromagnetic physics models (valid from 250 eV - 100 GeV) are used to model photon interactions with materials. 
The detector energy resolution and experimental threshold values were incorporated properly in the simulation to reproduce the experimental spectra. The statistical errors in the 
simulated photopeak area have been considered in the simulated efficiency plot. 

\begin{figure}
\centering
\includegraphics[height=8.5cm,width=10.5cm]{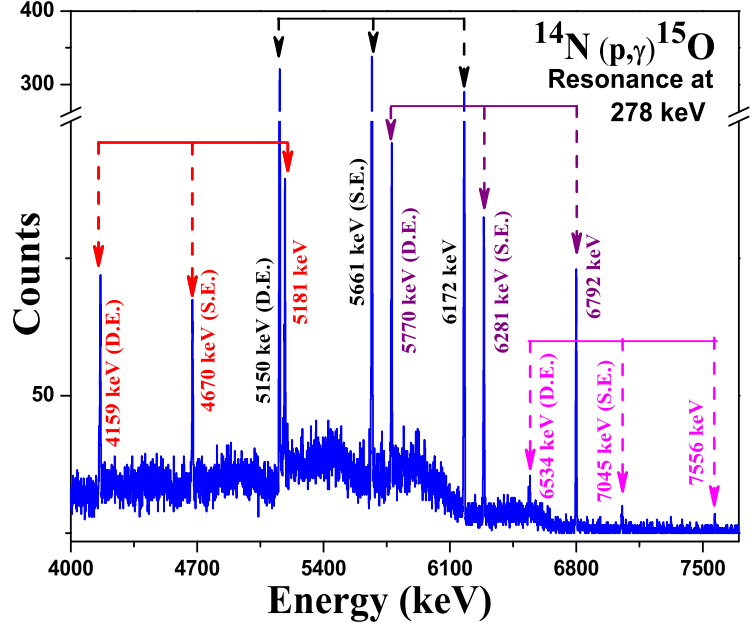} 
 \caption{
\label{fig4_inbeamspec}
Typical high-energy part of $\gamma$-ray spectra from $^{14}$N(p,$\gamma$)$^{15}$O resonance reaction at E$_{p}^{lab}$= 278 keV. S.E. and D.E. denote single escape and double escape peaks, respectively.  }
\end{figure}

\begin{figure}
\begin{center}
\includegraphics[width=\textwidth]{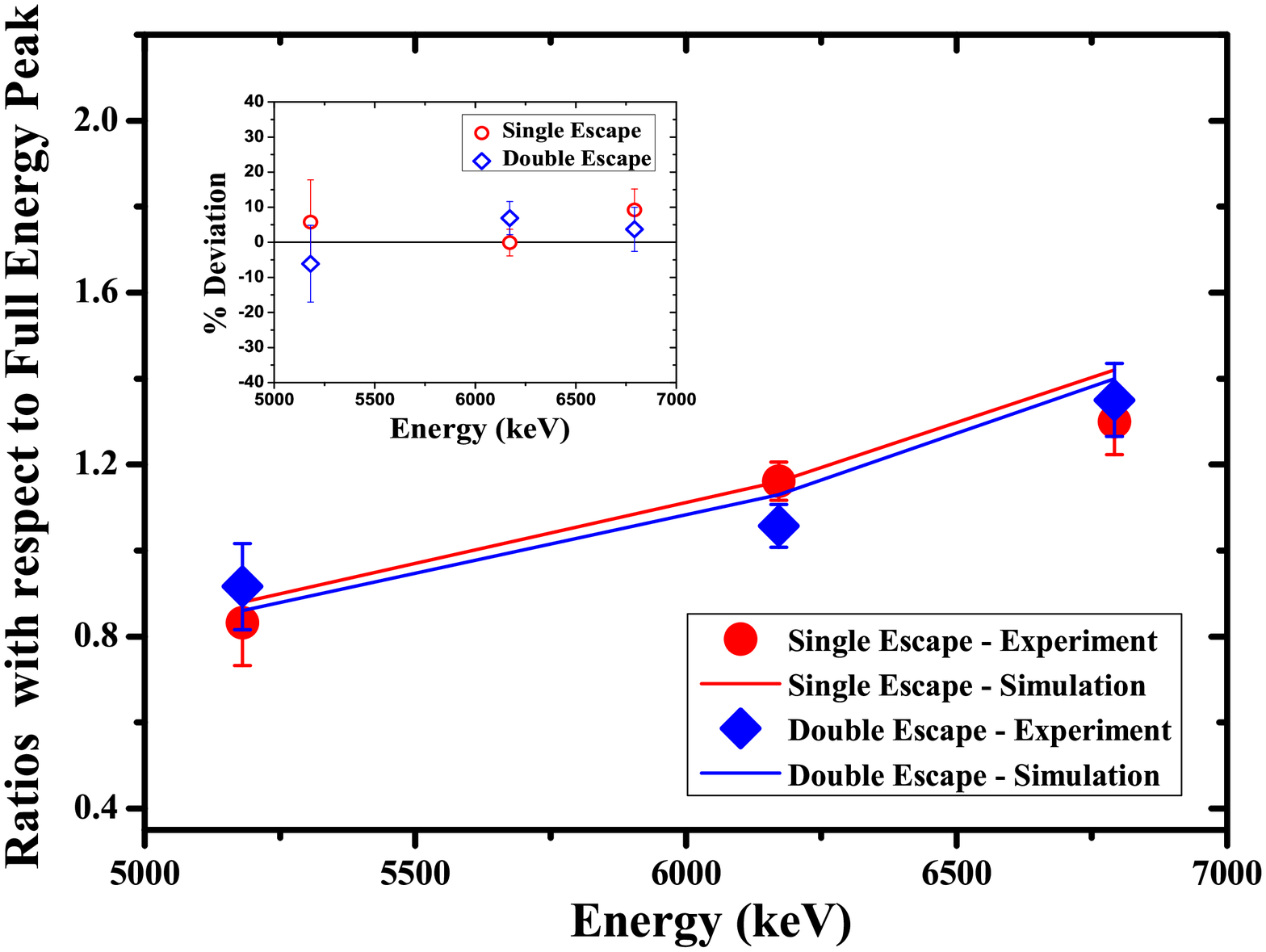}
\caption{\label{fig5_ratio} \small \sl The relative contribution for S.E. and D.E. with respect to F.E. as a function of $\gamma$- energy for the 
BEGe detector. The percentage deviation of the experimental value from simulation is shown in the in-set figure.}
\end{center}
\end{figure}

\subsubsection{The detector geometry}

The FALCON BEGe  detector was simulated with the geometry and specifications provided by the manufacturer. The necessary details 
about it and the internal structure are provided  in Fig. \ref{fig1_falcon},  and  Table \ref{detector}.

\subsubsection{The source holder and beam-pipe geometry }
For in-beam efficiency calibration from the resonance reaction data, detailed information regarding the  source holder, the clamp to bind it to the beam-pipe and the part of the beam-pipe through which the gammas have to travel to reach the detector were included
in detail in the simulation. The details are  provided in Fig. \ref{fig1_falcon} and in Table \ref{holder}.

\section{Results and Discussions}
\subsection{Experimental results}
\subsubsection{Energy calibration }
The detector was calibrated with the laboratory - standard $\gamma$- ray sources and in-beam spectra of resonance reaction.  The $\gamma$- rays emitted from a $^{152}$Eu radioactive source and the resonant level in $^{15}$O (7.556 MeV) \cite{iliadis} were used to calibrate the detector upto 7 MeV. The deviations from linear and quadratic fits to the calibration curves for the Falcon detector from 0.122 MeV to 7 MeV is shown in Fig. \ref{fig2_calib}. It is evident from the figure that quadratic fit is essential for energies above 3 MeV. 

\subsubsection{Energy resolution}

The energy resolution of the BEGe detector at different energies is shown in Fig. \ref{fig3_fwhm}. The resolution is defined as, full width at half maximum (FWHM) of the peak ($\Delta$E). The resolutions are plotted as function of the  square root of energy. The plots for radioactive source and the in-beam spectra at 0$^o$ are fitted with second order and third order polynomials, respectively.
It is evident from the figure as well as the inset table  that 
the energy dependences of resolution for the radioactive source $^{152}$Eu (Set I) and in-beam 
$\gamma$- rays emitted by the decay of the resonance state (Set II) are totally different. It has been noticed that this dependence also varies as a function of the angle of the detector with respect to target  position.
This can be explained if the lifetimes of the states corresponding to the de-excitation $\gamma$- rays are considered. The resonance state at 7.556 MeV has  T$_{1/2}$ =0.99 (10) keV, with  
T$_{1/2}$  of 5.181, 6.172 and 6.792 MeV states reported as 5.7 (7) fs, $<$ 1.74 fs and  $<$ 20 fs, respectively \cite{ensdf}. 
The 90$^o$ in-beam spectra show maximum broadening for the peaks, whereas the 0$^o$ peaks  are shifted, but they have minimum broadening. The figure shows the energy dependence of the 0$^o$ spectra.
The finite opening angle of the detector and the proximity of the detector to the target also contributes to the peak broadening in the in-beam spectra. 
The FWHM at 1.408 keV of $^{152}$Eu is 2.17$\pm$0.02 keV, whereas that at around 7 MeV for in-beam spectra is 4.72$\pm$0.17 keV. These energy dependences have been incorporated while simulating  the $^{152}$Eu and in-beam spectra.

\begin{figure}
\begin{center}
\includegraphics[height=12.0cm,width=16.0cm]{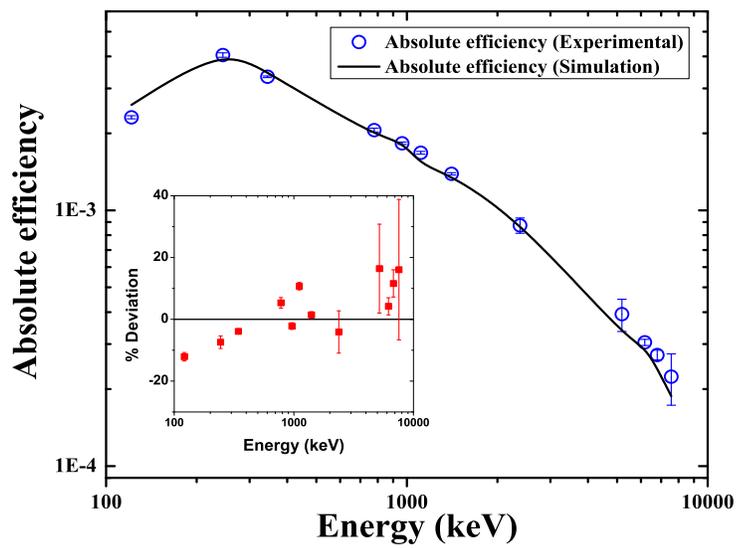} %{fig9_abseffi.eps}
\vspace{-3cm}
\caption{\label{fig6_abseffi} \small \sl Absolute detection efficiency as a function of energy from 0.122 MeV to 7 MeV. Open blue circles represent the experimental data for Falcon BEGe at 50$^{0}$. The corresponding error bars have been included in the symbols. The simulated results are shown with a solid line. The inset shows the percent deviation of the experimental data from simulated results. The errors in the deviation plot is only from the data. }
\end{center}
\end{figure}

\begin{figure}
\begin{center}
\includegraphics[width=\textwidth]{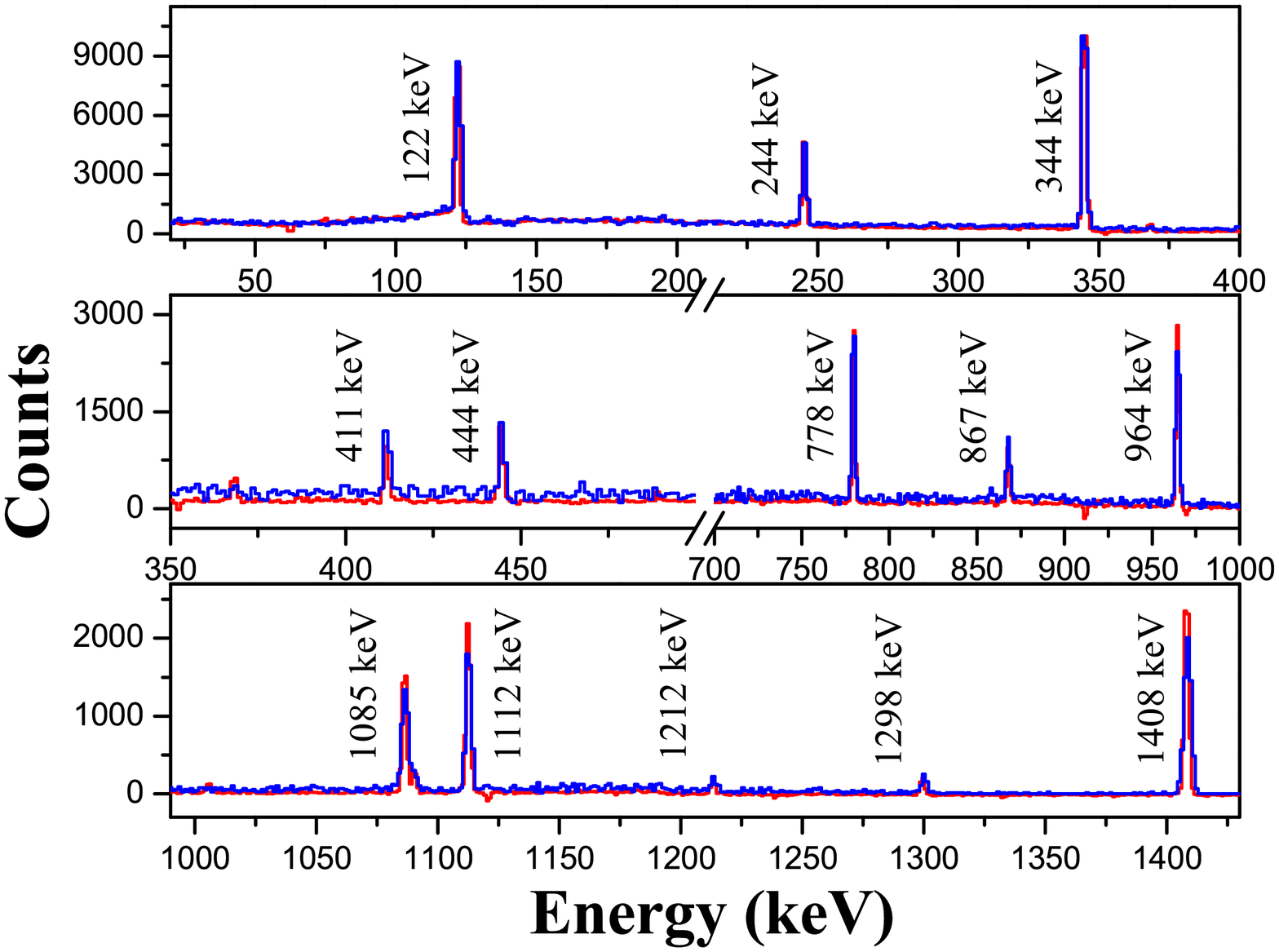}
\caption{\label{fig7_152Eusimu} \small \sl Comparison of experimental and simulated spectra at an angle 50$^{0}$ of $^{152}$Eu for the BEGe detector. The experimental and simulated spectra are drawn with red and blue curves, respectively. }
\end{center}
\end{figure}

\begin{figure}
\begin{center}
\includegraphics[width=8.0cm,width=\textwidth]{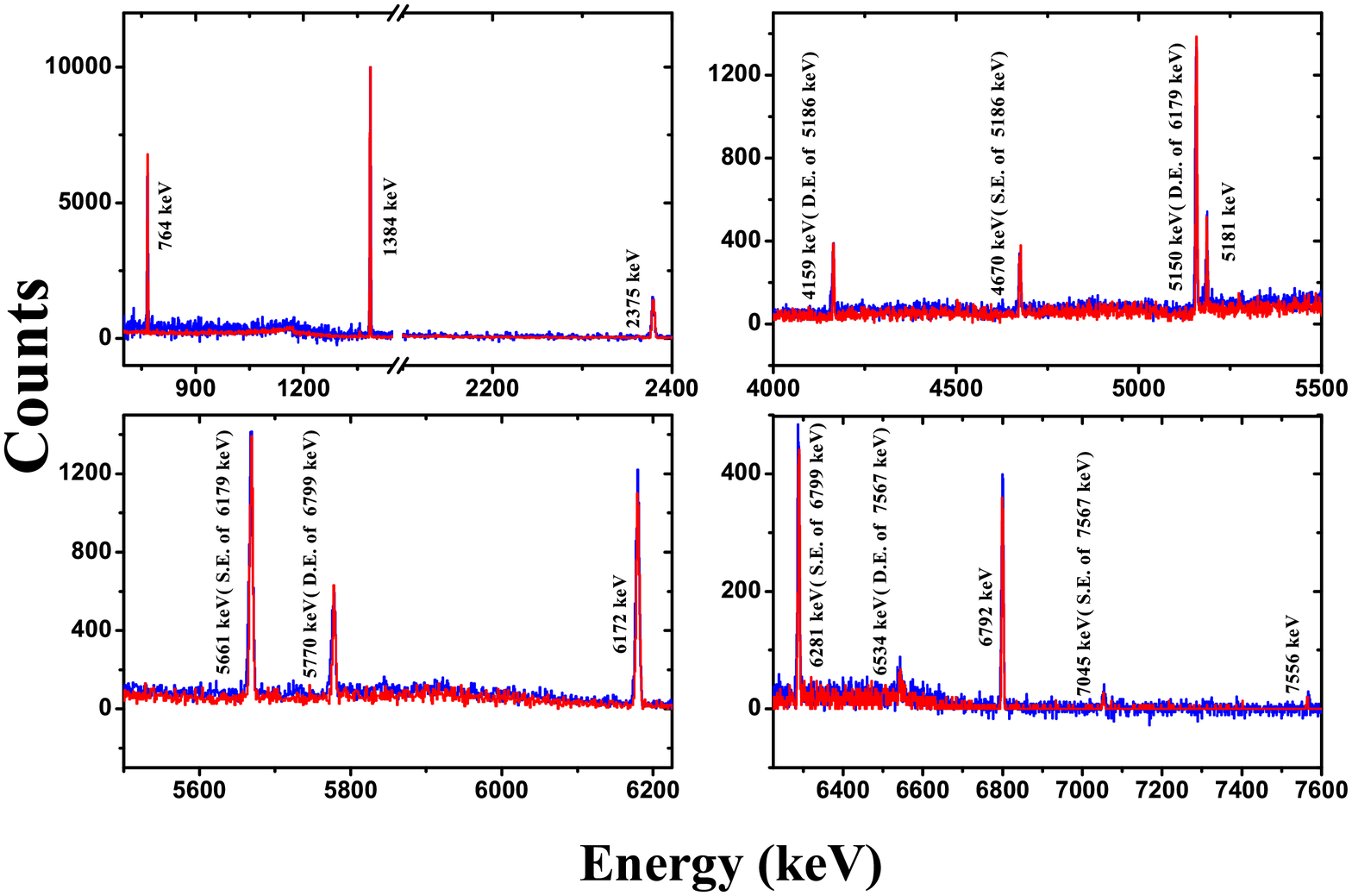}
\caption{\label{fig8_inbeamsimu} \small \sl Comparison of experimental and simulated in-beam spectra at an angle 50$^o$ for the BEGe detector. The experimental and simulated spectra are drawn with red and blue curves, respectively.}
\end{center}
\end{figure} 

\begin{figure}
\begin{center}
\includegraphics[width=1.5\textwidth]{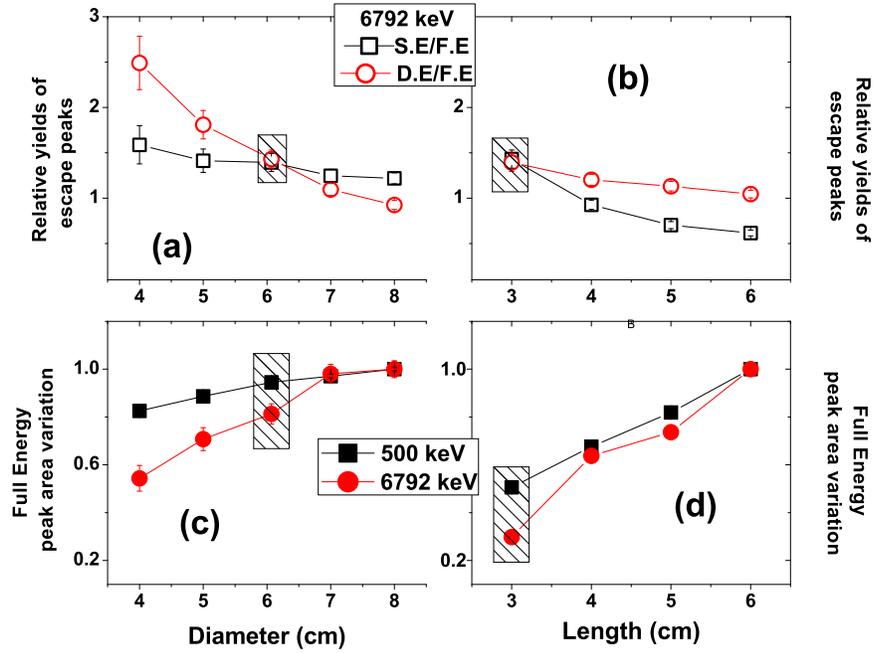}
\vspace{-3cm}
\caption{\label{fig9_ratio_simu} \small \sl The study of the effect of variation of crystal diameter and length on escape peaks and full energy 
peak by simulation. The relative yields of escape peaks for 6792 keV $\gamma$- ray are shown with variation of (a) diameter and (b) length of the crystal.
The variation in the full energy peak yields for 500 keV and 6792 keV $\gamma$- rays are shown with increase in (c) diameter, with areas normalised by 
radius$^2$ to eliminate the effect of solid angle variation and (d) length of the crystal. The shaded boxes indicate the features corresponding to the dimensions of the present detector.}
\end{center}
\end{figure}

\begin{figure}
\begin{center}
\includegraphics[width=\textwidth]{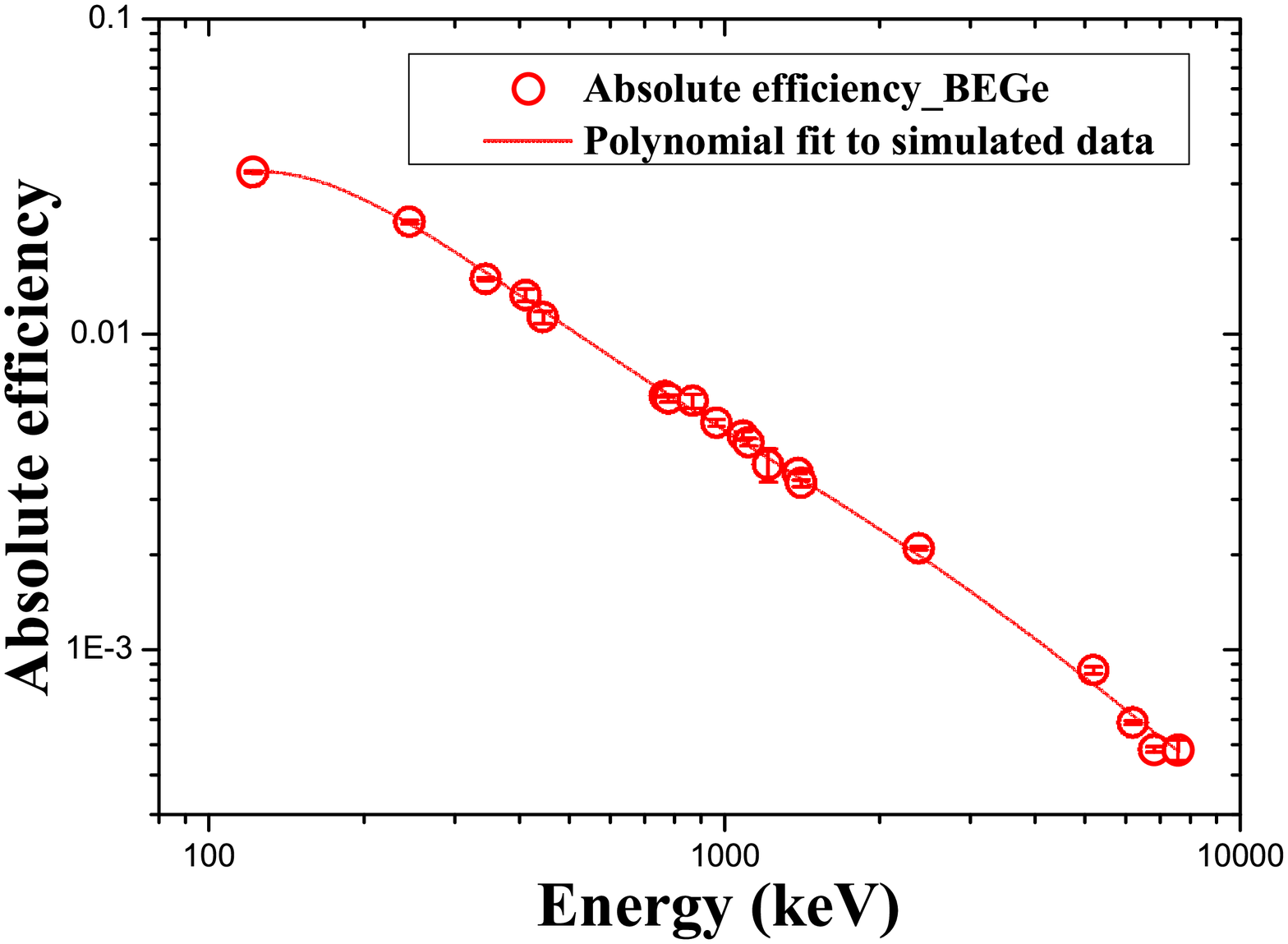}
\caption{\label{fig10_abssimu} \small \sl Simulated absolute detection efficiency values for $^{152}$Eu and resonance reaction $\gamma$- ray energies. Open red circles represent the results for Falcon BEGe detector. The corresponding error-bars 
are statistical errors calculated from the simulated photo-peak areas.}
\end{center}
\end{figure}

\subsubsection{The escape peaks}

The high energy portion of typical $\gamma$- ray spectra from the detector for gammas emitted from the resonance state are shown in 
Fig. \ref{fig4_inbeamspec}.  The single (S.E.) and double escape (D.E.) peaks are marked in the figure. The relative intensities of the escape peaks, vis-a-vis the full energy peak (F.E.) clearly show an energy dependence.  The experimental relative ratios of the escape peaks with respect to full energy peak is shown in Fig. \ref{fig5_ratio}.  The  higher intensity of the double escape peak among the three peaks (D.E., S.E. and F.E.) has been observed in the BEGe beyond 5 MeV. 
The simulation results  reliably reproduce the experimental trends.
%The percentage deviation of the experimental value from simulation is shown in the in-set figure.

\subsubsection{Absolute efficiency measurement}

The absolute efficiency at low energy region was measured using $^{152}$Eu source with known strength, whereas the intensities of $\gamma$- rays from the resonant levels in $^{15}$O (7.556 MeV) \cite{iliadis} were used for efficiency estimation in the higher energy region. 
To get the experimental absolute efficiency at higher energy, we have normalized the efficiency data from $^{152}$Eu and resonance reaction at 1.408 MeV. The absolute efficiencies of Falcon detector at 50$^{o}$ are shown  (see Fig. \ref{fig6_abseffi}), to minimize the
angular distribution effect for the in-beam resonance reaction data. The attenuation of the $\gamma$- ray at the target chamber is evident from the low energy part of the curve. The absolute efficiencies at 6.7 and 7.5 MeV are 2.72e-04$\pm$1.19e-05 and 2.23e-04$\pm$5.08e-05, respectively.
%6799.6	2.72089E-4	2.40502E-4
%7567.19	2.23789E-4	1.87821E-4

\subsection{Comparison of Simulation results with Experimental data}
\subsubsection{The spectra}
The $\gamma$- ray spectra generated from GEANT4 simulation with 
$^{152}$Eu $\gamma$- ray  intensities taken from ENSDF data library \cite{ensdf} for BEGe detector are shown in Fig. \ref{fig7_152Eusimu}.  
The room background - subtracted  experimental $\gamma$- ray spectra of $^{152}$Eu have also been plotted  in Fig. \ref{fig7_152Eusimu} for comparison. The energy dependence of the resolution
has been incorporated from the fitting parameters (Set I) of the  plots shown in Fig. \ref{fig3_fwhm}. 

The simulations for $\gamma$- rays from the resonant level of $^{15}$O for BEGe detector are displayed in Fig. \ref{fig8_inbeamsimu}. The $\gamma$- rays emitted by  the resonance state 
7.556 MeV of $^{15}$O were thrown randomly with proper intensity to get the simulated spectra. 
The BEGe in-beam resonance spectra at 50$^o$ (to minimize angular distribution effect) match reasonably well with the simulated one. The energy dependence of the resolution
has been incorporated from Set II. The statistical consistency check is performed on a point-by-point basis for the data points in Figs. \ref{fig7_152Eusimu} and \ref{fig8_inbeamsimu}. The simulated points are always within 1$\sigma$ level from the experimental points. 

\subsubsection{The effect of variation in the crystal dimensions}

The dependence of the
ratios of yields of single escape and double escape peaks for 6792 keV $\gamma$- rays with respect to the full energy yields  on variations
in the diameter and length of the BEGe crystal obtained from simulation are shown in Fig. \ref{fig9_ratio_simu}. From Fig. \ref{fig9_ratio_simu}a, it 
is evident that the increase in 
diameter reduces the double escape peaks of 6792 keV $\gamma$- ray with respect to the full energy peaks faster than the decrease of single escape peaks. Single escape peaks appear to remain almost the same. A part of these events  decreases, increasing the F.E. peak. However, S.E. peak also  increases due to reduction of D.E. peaks. On the other hand,  the increase in length reduces the S.E. peaks with respect to the full energy peaks faster than the decrease of double escape peaks, indicating that the full energy detection efficiency improves faster with increase in length of the crystal (Fig. \ref{fig9_ratio_simu}d). The variations in the yields of full energy peaks of 500 keV and 6792 keV $\gamma$- rays  with increase in diameter and length of the crystal are also simulated and shown in Fig. \ref{fig9_ratio_simu}c and \ref{fig9_ratio_simu}d. For diameter changes, peak areas / radius$^2$ of the crystal
have been compared to eliminate the effect of solid angle variation in the yield. Increase in length and diameter has more significant improvements in the
detection efficiencies of 6792 keV gamma compared to 500 keV.

\subsubsection{The efficiency for a bare point source}

The efficiency of the detector has been determined from simulated spectra of $^{152}$Eu and in-beam, to generate the simulated efficiency curve 
for the BEGe detector in the present setup (Fig. \ref{fig6_abseffi}). The  percent deviations between experimental efficiencies
and the simulated values calculated using the following formula are shown in the inset of the figure. 
\begin{equation}
\% Deviation = (eff_{exp}-eff_{simu})/eff_{exp}*100
\end{equation}

Later, with the same detector configuration, absolute efficiencies of the detector have been determined for bare point sources at a distance of 5 cm from the detector front face without any target chamber in between. The simulated results are shown in Fig. \ref{fig10_abssimu}. The low energy part of the curve now shows the characteristics of a BEGe crystal
in contrast to the strong attenuation effect seen in the experimental data (see Fig. \ref{fig6_abseffi}). The simulated efficiency at 7 MeV is around 4.8e-04$\pm$3.59e-05.
Thus, these electrically cooled BEGe detector could be effectively used in the $\gamma$- ray spectroscopy as well as for the study of resonance phenomena in nuclear astrophysics.

\section{Summary and Conclusions}

In summary, an electrically cooled FALCON BEGe detector has been characterized using the low energy laboratory - standard $^{152}$Eu source.  The characterization has been extended to higher energies (till 7 MeV) by utilizing an in-beam experiment to measure the high energy $\gamma$- rays (up to 7 MeV) decaying from a resonance state of $^{15}$O populated in the resonance reaction 
$^{14}$N(p,$\gamma$) at E$_{p}^{lab}$= 278 keV. The experimentally determined characteristics are well reproduced by the GEANT4 simulation.  
It has been found that the electrically cooled BEGe detectors can be efficiently used in high-resolution in-beam $\gamma$- ray spectroscopy 
along with its intended use in the measurement of high energy $\gamma$- rays for studying the resonance phenomena in low energy nuclear astrophysics.

\section{Acknowledgements}
%\acknowledgments

We would like to thank all the members of ECRIA lab at TIFR, Mumbai for their help and cooperation. Special thanks are due to Mr. Suraj Kumar Karan
from SINP for his technical help in AutoCAD drawings.

\end{document}